\documentclass[
reprint,
superscriptaddress,
nofootinbib,
nobibnotes,
amsmath,amssymb,
aps,
prl,
]{revtex4-2}

\usepackage{graphicx}
\usepackage{dcolumn}
\usepackage{bm}
\usepackage{lipsum}
\usepackage{float}
\usepackage{xcolor}
\usepackage{soul}
\usepackage{tabularx}
\usepackage{makecell}
\usepackage{amssymb}
\usepackage[normalem]{ulem}
\usepackage{braket}
 \usepackage{hyperref}
    \hypersetup{
        pdftitle={Orbitally tuned composite-fermion pairing transitions in bilayer graphene},
        colorlinks=true,
        linkcolor=blue,
        anchorcolor=blue,
        citecolor=blue,
        urlcolor=blue,
    }

\newcommand{\vect}[1]{{\bf #1}}

\begin{document}

\title{Orbitally tuned composite-fermion pairing transitions in bilayer graphene}
\author{Ravi Kumar}
\altaffiliation{These authors contributed equally to this work.}
\affiliation{Department of Condensed Matter Physics, Weizmann Institute of Science, Rehovot, 76100, Israel.}
\author{Tomer Firon}
\altaffiliation{These authors contributed equally to this work.}
\affiliation{Department of Condensed Matter Physics, Weizmann Institute of Science, Rehovot, 76100, Israel.}
\author{André Haug}
\altaffiliation{These authors contributed equally to this work.}
\affiliation{Department of Condensed Matter Physics, Weizmann Institute of Science, Rehovot, 76100, Israel.}
\author{Misha Yutushui}
\affiliation{Department of Condensed Matter Physics, Weizmann Institute of Science, Rehovot, 76100, Israel.}
\author{Alon Ner Gaon}
\affiliation{Department of Condensed Matter Physics, Weizmann Institute of Science, Rehovot, 76100, Israel.}
\author{Kenji Watanabe}
\affiliation{Research Center for Functional Materials, National Institute for Materials Science, 1-1 Namiki, Tsukuba 305-0044, Japan.}
\author{Takashi Taniguchi}
\affiliation{International Center for Materials Nanoarchitectonics, National Institute for Materials Science, 1-1 Namiki, Tsukuba 305-0044, Japan.}
\author{David F. Mross}
\affiliation{Department of Condensed Matter Physics, Weizmann Institute of Science, Rehovot, 76100, Israel.}
\author{Yuval Ronen}
\email{yuval.ronen@weizmann.ac.il}
\affiliation{Department of Condensed Matter Physics, Weizmann Institute of Science, Rehovot, 76100, Israel.}
\date{\today}
\begin{abstract}
The effective interaction between composite fermions, set entirely by the Coulomb potential and the underlying electronic Landau level orbitals, can stabilize exotic fractional quantum Hall states. In particular, half-filled Landau levels with different orbital character can host either metallic or paired phases of composite fermions. Here, we leverage experimental control over the orbital composition to realize a composite-fermion pairing transition in the first excited Landau level of bilayer graphene. Transport measurements at filling factors $\nu=\frac{9}{2}$ and $\frac{11}{2}$ reveal conductive states giving way to well-developed plateaus with increasing displacement fields. These states are insensitive to an in-plane magnetic field, indicating single-component ground states and thus pointing at non-Abelian orders. Our numerical study, based on displacement-field-dependent Landau-level wavefunctions, supports the orbital origin of the pairing transition and suggests Moore--Read or anti-Pfaffian ground states.
\end{abstract}
\maketitle
\textit{Introduction}---Electron--electron interactions are promoted to a zeroth-order effect in the fractional quantum Hall (FQH) regime, where the kinetic energy is quenched. Still, the resulting phases are sensitive to the electronic single-particle states via orbital form factors, a property of their kinetic Hamiltonian. As such, the distinct orbital wavefunctions in different Landau levels (LLs) can stabilize competing ground states, ranging from compressible states to gapped quantum Hall liquids and electronic crystals \cite{Chakraborty_1995,pinczuk1997perspectives,fogler2002stripe, Jain_composite_2007,fradkin2010nematic}.

The half-filled lowest LL ($N=0$) in narrow GaAs quantum wells typically hosts a compressible composite-Fermi liquid (CFL) \cite{halperin_theory_1993}. In contrast, the first excited one ($N=1$) forms an incompressible state with a quantized plateau in the Hall conductance $\sigma_{xy}$~\cite{Willett_observation_1987}, which is  interpreted as composite-fermion pairing; see Ref.~\onlinecite{kunyangreview} for a recent review. Analogous behavior at half filling, realizing either CFL or incompressible even-denominator states, has also been observed in alternative GaAs quantum well structures~\cite{Suen_Observation_1992,Eisenstein1992DoubleLayer, Liu_2014_FQH_hole-side}, as well as in ZnO \cite{Falson_Zno_2015,Falson_Zno_2018,Falson_Phase_2019}, WSe\textsubscript{2} \cite{Shi_even_wse2_2020}, monolayer graphene \cite{Zibrov_Even_Denominator_2018, Kim_Even_Denominator_f_wave_2019}, and multilayer graphene \cite{Ki_bilyaer_graphene_2014,Kim_bilayer_graphene_2015,Li_bilayer_graphene_2017,Zibrov_Tunable_bilayer_graphene_2017, Huang_Valley_bilayer_graphene_2022,kumar_quarter_2025,chanda2025even,Hu2025,haug2026interaction}. The competition between compressible and incompressible states~\cite{Morf_transition_1998,Park_possibility_1998,Rezayi_incompressible_2000,Scarola_Cooper_2000,Wojs_3body_2005,feiguin_density_2008,Peterson_Orbital_2008,moller_paired_2008,Feiguin_spin_2009,Zaletel_Topological_2013,Sharma_BCS_2021,Sharma_CF_pairing_quart_2024,Yutushui_phase_2025} is attributed primarily to the specific LL wavefunctions, which modify the electron--electron repulsion at short distances~\cite{Haldane_fqh_1983}. Additional factors that could promote pairing include screening of the Coulomb potential and LL mixing \cite{Zhao_composite_2023}.

To better understand the conditions that lead to composite-fermion (CF) pairing, a platform whose LL orbitals can be tuned efficiently is beneficial. Experimental control over the single-particle states can be achieved straightforwardly through perturbations that mix different bands. In bilayer graphene (BLG), the displacement field $D$ can play this role. Its effect on the LL wavefunctions is set by the coupling strength relative to the energy splitting between the relevant bands. While in the zeroth LL, this energy difference is set by the atomic interlayer tunneling, in the first excited LL, it is given by the much smaller cyclotron frequency. Consequently, the first excited LL of BLG offers ideal conditions for tuning interactions between CFs via the orbital composition of the electronic LL wavefunctions.

In this work, we demonstrate orbitally-tuned transitions between compressible and incompressible FQH states in BLG at filling factors $\nu=\frac{9}{2}$ and $\nu=\frac{11}{2}$. We observe that the half-filled states are compressible at small $|D|$, while an activation gap gradually develops below $D_\mathrm{c} \approx -540~\mathrm{mV/nm}$. We attribute this gap to CF pairing, which sets in when the attractive interaction strength is tuned beyond a critical threshold. The $D_\mathrm{c}$ remains similar across different samples and cooldowns, see SI Secs.~3 and 6. The occurrence of these paired states at large displacement fields, together with their insensitivity to an in-plane magnetic field, points to single-component ground states, compatible with non-Abelian FQH phases. However, more direct measurements, such as thermal conductance or interferometry, are required to confirm the non-Abelian order.
\begin{figure*}
\centering
\includegraphics[width=1\textwidth]{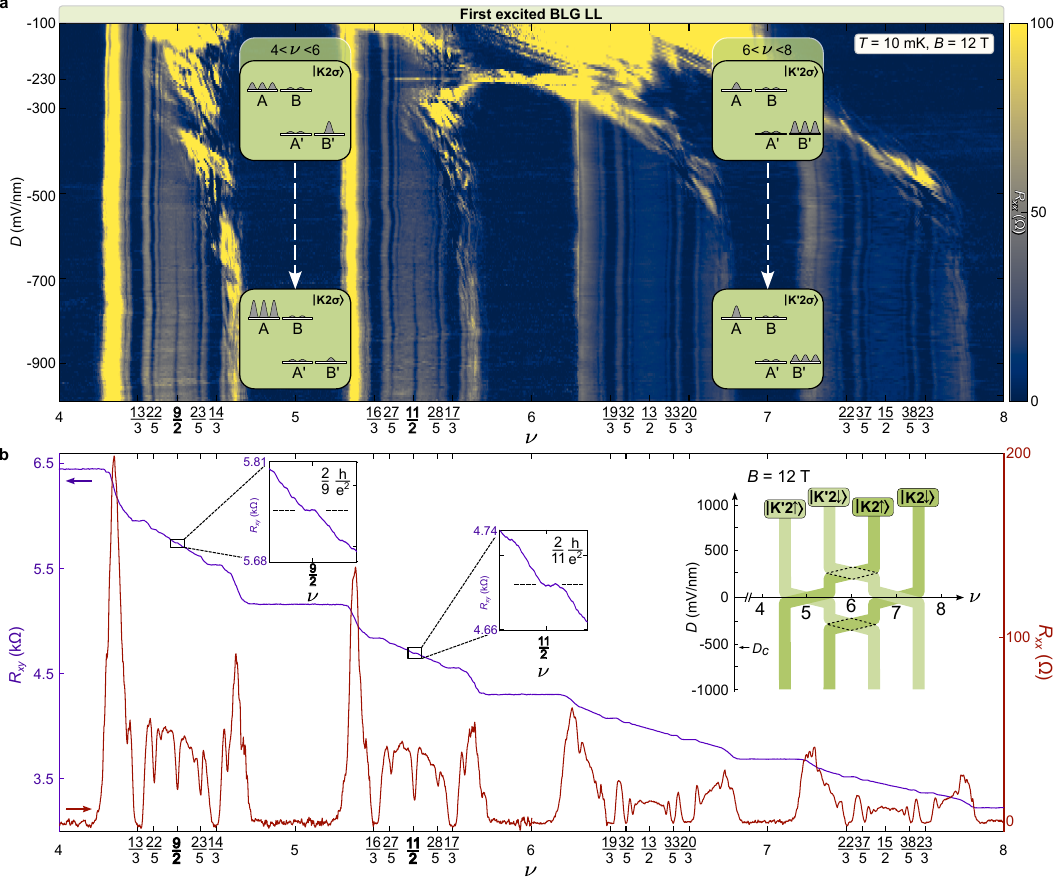}
\caption{\textbf{FQH states in the first excited LL of BLG}. \textbf{(a)} $R_{xx}$ as a function of $\nu$ and $D$, taken at $T = 10~\mathrm{mK}$ and $B = 12~\mathrm{T}$. The insets illustrate how the displacement field tunes the single-electron wavefunction composition across the BLG sites. The notation $\ket{\mathrm{K}2\sigma}$ or $\ket{\mathrm{K'}2\sigma}$ denotes a state with valley quantum number K or K', LL index 2 (the first excited BLG LL), and spin quantum number $\sigma$ ($\uparrow$ or $\downarrow$). At increasing negative $D$, the weight shifts to the $A$ sites, which are occupied by $N=2$ orbitals for $4\leq\nu\leq 6$, and by $N=0$ orbitals for $6\leq\nu\leq8$. Half-filled FQH states emerge below $-540~\mathrm{mV/nm}$ at $\nu=\frac{9}{2}$ and $\frac{11}{2}$, but not at $\nu=\frac{13}{2},\frac{15}{2}$. The changes in orbital weights shown in the inset are intended only as a qualitative illustration. \textbf{(b)} Longitudinal $R_{xx}$ and transverse $R_{xy}$ resistances as a function of $\nu$, averaged over the data measured in $D= -950~\mathrm{to}~-980~\mathrm{mV/nm}$ range. The $R_{xx}$ dips at $\nu=\frac{9}{2}$ and $\frac{11}{2}$ are accompanied by $R_{xy}$ plateaus, shown in the insets. The inset on the right illustrates crossings of spin- and valley-resolved levels inside the first excited LL of BLG. The $D$-axis values are added to qualitatively represent the experiment values in panel (a). The dashed black diamond boxes present the LL crossings at $D=\pm~230~\mathrm{mV/nm}$. }
\label{fig:2D_plot}
\end{figure*}

To understand the origin of this behavior, we study LL wavefunctions obtained from the four-band model of BLG \cite{McCann_BLG_2006, McCann_BLG_2013, Jung_Accurate_BLG_2014, hunt_direct_2017}. The first excited LL in a four-band model is composed of the $N=0$, $1$, and $2$ orbitals, whose relative weights depend sensitively on $D$, see insets in Fig.~\ref{fig:2D_plot}a. Increasing $|D|$ enhances the $N=2$ components at $\nu=\frac{9}{2}$ and $\frac{11}{2}$, thereby reducing short-range repulsion and favoring CF pairing at intermediate orbital mixing. Consistent with the experimental onset of incompressibility, our exact-diagonalization study finds that the half-filled ground state for $|D| \gtrsim 500~\mathrm{mV/nm}$ exhibits a large overlap with the Moore–Read Pfaffian \cite{Moore_nonabelions_1991} (or anti-Pfaffian \cite{Lee_particle_hole_2007,Levin_particle_hole_2007}) state.

\textit{FQH states in the first excited LL of BLG}---We fabricated two heterostructures that consist of BLG encapsulated by hexagonal boron nitride (hBN) and graphite layers acting as gates, using standard dry transfer technique~\cite{pizzocchero2016hot}. Longitudinal ($R_{xx}$) and transverse ($R_{xy}$) resistances were measured using a standard low-frequency lock-in technique; see Sec.~1 in \cite{SM} and also Refs.~\cite{kumar_quarter_2025,haug2026interaction} for details where same devices were studied. The results presented in the main text were obtained from one device; complementary data from this device and consistent data from a second device are shown in Sec.~3 in \cite{SM}. 

Fig.~\ref{fig:2D_plot}a shows $R_{xx}$ of the first excited LL as a function of $\nu$ and $D$ at a magnetic field of $B=12~\mathrm{T}$ and a base temperature of $T=10~\mathrm{mK}$. Dark-blue regions mark vanishing $R_{xx}$, indicating an incompressible bulk. The strongest states arise at the integer fillings $\nu=5,6,7$ with well-quantized plateaus in $R_{xy}$; see Fig.~\ref{fig:2D_plot}b. These integer quantum Hall states reflect the lifting of spin $\uparrow,\downarrow$ and valley $K,K'$ degeneracies. To identify the LL quantum numbers, we note that the Zeeman splitting dominates the energies at small $|D|$, while at larger displacement fields, the levels are ordered according to their valley. These two limits are separated by a level crossing, which we observe at $D \approx -230~\mathrm{mV/nm}$; see Sec.~5 in \cite{SM} for evidence of level crossing, consistent with previous observations \cite{Huang_Valley_bilayer_graphene_2022,kumar_quarter_2025,maryenko2014polarization, kim2023robust, liu2011evolution}. 

At partial LL fillings, numerous FQH states arise. For large displacement fields we observe strong $R_{xx}$ minima at $\nu=\frac{9}{2}$ and $\frac{11}{2}$, consistent with recent local-probe measurements~\cite{Hu2025}. Our measurements show that both half-filled states evolve gradually out of compressible states below $D \lesssim -540~\mathrm{mV/nm}$ and exhibit emerging $R_{xy}$ plateaus. The plateaus emerge with $D$ over a range of $400~\mathrm{mV/nm}$. To demonstrate the quantization of the plateaus, in Fig.~\ref{fig:2D_plot}b, we plot $R_{xx}$ and $R_{xy}$ as a function of $\nu$, averaged over the data measured in the $D= -950~\mathrm{to}~-980~\mathrm{mV/nm}$ range. The measured $R_{xy}$ plateau values are within $2~\%$ of the expected values; see Sec.~3 in \cite{SM} and Table~\ref{tableddata} in the End Matter. As the half-filled states strengthen with $D$, the nearby Jain states weaken. 

The enhancement (suppression) of the $R_{xx}$ minima at even (odd) denominators is consistent with an increased attractive interaction between CFs. At half filling, the CFs form a paired superfluid whose gap grows with the strength of this interaction~\cite{Read_paired_2000,Scarola_Cooper_2000}. 
At Jain-state fillings where CFs form integer quantum Hall states, the attraction mixes the CF LLs, thus reducing the gap \cite{Yutushui_daughters_2024}. The continuous deformation of the electron orbitals, illustrated in the left inset of Fig.~\ref{fig:2D_plot}a, provides a mechanism for tuning the CF interaction. In particular, the weight of the $N=2$ orbitals increases with $|D|$, which reduces the electron--electron repulsion at short distances, presumably leading to attraction between CFs (see Sec.~6 in \cite{SM} and the end matter).

In contrast, the $R_{xx}$ measurements for $6<\nu<8$ do not exhibit any significant $D$ dependence. In particular, the fillings $\nu=\frac{13}{2}$ and $\nu=\frac{15}{2}$ are compressible throughout. This behavior is consistent with an increased $N=0$ weight, which enhances short-range repulsion between electrons and disfavors pairing of CFs. At very low $|D|$, we do not expect any significant changes in $R_{xx}$ in the entire first excited LL. However, the transport signatures are overshadowed by artifacts due to contact doping~\cite{kumar_quarter_2025}.

\textit{Evolution of activation gaps}---Next, we measure the activation gaps $\Delta$ of the half-filled and surrounding Jain states as a function of $D$. This is achieved by measuring the $R_{xx}$ minimum at elevated temperatures where we observed an activated behavior of $R_{xx}$, see Sec.~7 in \cite{SM} for details.
\begin{figure} 
    \includegraphics[width=0.48\textwidth]{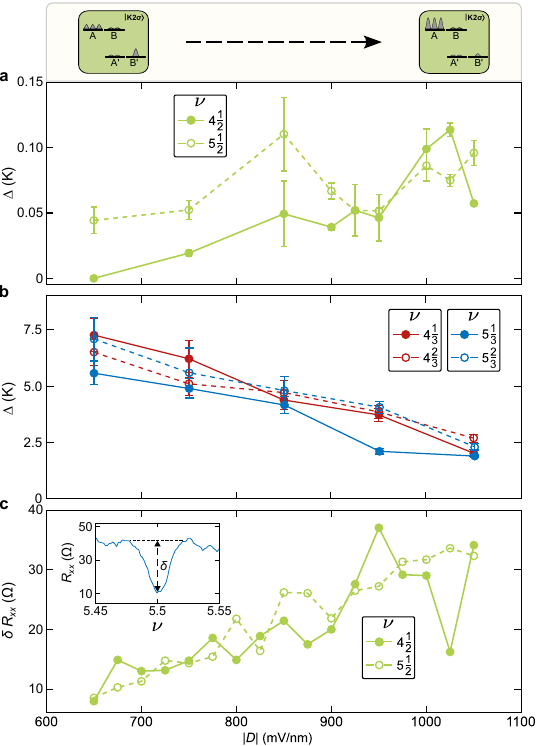}
    \caption{\textbf{Strength of paired and Jain FQH states as a function of $D$.} \textbf{(a)} The activation gaps $\Delta$ at $\nu=\frac{9}{2}$ and $\frac{11}{2}$ increase with negative $D$. The insets above illustrate the orbital composition of the LL wavefunctions for low and high displacement fields. \textbf{(b)} The gaps of Jain states at $\mathrm{4 +1/3}$, $\mathrm{4 +2/3}$, and $\mathrm{5 +1/3}$, $\mathrm{5 +2/3}$ fillings decrease over the displacement-field range where half-filled gaps increase. \textbf{(c)} The strengths of the resistance minima $\delta R_{xx}$ follow the same trend as the gaps in (a). $\delta R_{xx}$ was determined as illustrated in the inset for $|D|=975~\mathrm{mV/nm}$.}
    \label{fig:gapsVsD.pdf}
\end{figure}
In Fig.~\ref{fig:gapsVsD.pdf}a, we show the gaps of the half-filled states at $\nu=\frac{9}{2}$ and $\frac{11}{2}$, and those of the nearby Jain states at $\mathrm{4 +1/3}$, $\mathrm{4 +2/3}$, and $\mathrm{5 +1/3}$, $\mathrm{5 +2/3}$ fillings in Fig.~\ref{fig:gapsVsD.pdf}b. The data reveal an opposite trend of the gaps as a function of the displacement field: The half-filled gaps gradually grow with increasing $|D|$, while the Jain-state gaps shrink. The same trend is apparent in the strength of the $R_{xx}$ minima, shown for the half-filled states in Fig.~\ref{fig:gapsVsD.pdf}c. We note that the activation gaps of the half-filled states are small and are not intended to be presented as quantitative measurements, but rather as estimates that capture the overall trend. Note that the $R_{xx}$ data in Fig.~\ref{fig:2D_plot}a at these Jain states appears largely insensitive to $D$ due to their relatively large energy gaps, which strongly suppress thermal activation at $10~\mathrm{mK}$. At elevated temperatures, however, a clear and systematic $D$ dependence emerges, confirming that the activation gaps evolve with $D$ (see Secs.~9 and 10 in \cite{SM} ). A similar evolution with $|D|$ is observed for the gaps of the $\mathrm{4+2/5}$ and $\mathrm{5+2/5}$ states, as shown in Fig.~S10e in \cite{SM}.

The gradual onset of the half-filled gaps is markedly different from the behavior in BLG's zeroth LL. There, a small change in $D$ ($\approx10\mathrm{mV/nm}$) is sufficient to abruptly change the orbital character of the LL, switching from $N=0$ (no half-filled plateaus) to $N=1$ (well-developed half-filled states) at several level crossings \cite{Ki_bilyaer_graphene_2014,Kim_bilayer_graphene_2015,Li_bilayer_graphene_2017,Zibrov_Tunable_bilayer_graphene_2017,Huang_Valley_bilayer_graphene_2022, kumar_quarter_2025}. In contrast, the level crossing in the first excited LL appears to be unrelated to the development of a half-filled gap. These half-filled states evolve over a broader range of $D$ (several 100~$\mathrm{mV/nm})$.

\begin{figure}
\includegraphics[width=0.48\textwidth]{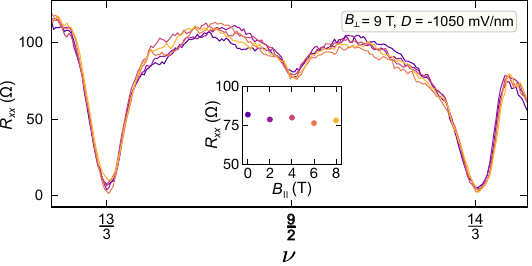}
    \caption{\label{fig:inplaneB.pdf} \textbf{Spin polarization of half-filled states}. $R_{xx}$ measurements around $\nu=\frac{9}{2}$ at $T=10~\mathrm{mK}$ and $B_\perp=9~\mathrm{T}$
    for different in-plane magnetic fields $B_\parallel$. The inset shows $\delta R_{xx}$ at $\nu=\frac{9}{2}$ as a function of $B_\parallel$.}
\end{figure}

\textit{Single-component nature of the paired states}---The occurrence of the half-filled states at very large $|D|$ suggests that they are fully valley-polarized.
To also check for their spin polarization, we measured the dependence of the $R_{xx}$ minima on in-plane magnetic fields $B_\parallel$ up to $8~\mathrm{T}$ at a constant out-of-plane magnetic field of $B_\perp=9~\mathrm{T}$. No appreciable change in $R_{xx}$  occurs, as shown for $\nu=\frac{9}{2}$ in Fig.~\ref{fig:inplaneB.pdf} and for $\nu=\frac{11}{2}$ in Sec.~4 in \cite{SM}.

The Zeeman energy at the largest in-plane field, around $10~\mathrm{K}$, is over two orders of magnitude larger than the half-filled gap, $\sim 50-100~\mathrm{mK}$ at $B_\perp = 12~\mathrm{T}$. We therefore conclude that the $\nu=\frac{9}{2}$ and $\frac{11}{2}$ states are fully spin-polarized, i.e., they realize a single-component wavefunction. Note that, to fully validate the spin polarization, a direct approach, such as nuclear magnetic resonance study is required \cite{tiemann2012unraveling,stern2012nmr}.

Next, we examined whether these states could correspond to competing broken-symmetry phases known to occur in higher Landau levels. In the range $4<\nu<6$, stripe and bubble phases would manifest as re-entrant integer quantum Hall phases which yield vanishing $R_{xx}$ with integer-quantized $R_{xy}$. In contrast, at $\nu=9/2$ and $\nu=11/2$, we observe clear minima in $R_{xx}$ accompanied by half-integer quantized fractional $R_{xy}$ plateaus, with no evidence of re-entrant behavior. Measurements with orthogonal current directions reveal a resistance anisotropy, $R_{xx}/R_{yy}\approx2$, much smaller than that reported for nematic fractional quantum Hall phases in GaAs\cite{wang2025even}, see sec.~12 in \cite{SM}. These results rule out any charge-ordered or strong nematic phases. Moreover, the nonlinear $I$–$V$ characteristics support the presence of incompressibility at $\nu=9/2$ and $\nu=11/2$, see sec~13 in \cite{SM}.

\textit{Orbital composition of BLG LLs}---To illustrate how the orbital compositions of the LLs evolve, we focus on the reduced two-band Hamiltonian for the A and B$'$ sites. The weight of the wavefunctions on the A$'$ and B sites is strongly suppressed by the atomic interlayer tunneling \cite{McCann_BLG_2006, McCann_BLG_2013, Jung_Accurate_BLG_2014, hunt_direct_2017}; including them does not significantly affect how the orbitals evolve with the displacement field (see Sec.~6 in \cite{SM} for the full four-band calculation).

The effective two-band Hamiltonian describing a single valley is $H=H_0+H_D$ with
\begin{align}
    H_0 = -\frac{1}{2m}\begin{pmatrix}
        0 & (\pi^\dagger)^2 \\ \pi^2 & 0
    \end{pmatrix}, \quad 
    H_D=\xi\frac{u}{2}\begin{pmatrix}
        1 & 0 \\ 0 & -1
    \end{pmatrix},
\end{align}
where $\xi=\pm 1$ for the $K$ or $K'$ valley and $u \propto D$ is the interlayer potential difference.  $H_0$ has two zero-energy eigenstates, $(\phi_0,0)$ and $(\phi_1,0)$, which form the zeroth LL of BLG. They are also eigenstates of $H_D$ and their orbital composition is thus unaffected by the displacement field.

\begin{figure}[t]
\includegraphics[width=\linewidth]{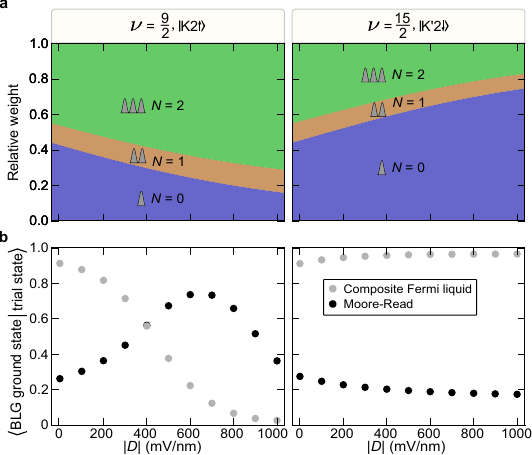}
\caption{ \textbf{Orbital composition and many-body ground states in the first excited LL of BLG.}
\label{fig:Theory2alt} \textbf{(a)} The single-electron states in BLG's first excited LL are composed of $N=0,1,2$ orbital wavefunctions. At $\nu=\frac{9}{2}$, a negative $D$ enhances the $N=2$ component at the expense of $N=0$, and vice versa at $\frac{15}{2}$. \textbf{(b)} The squared overlap between the Coulomb ground state and trial wavefunctions. The ground state changes from a CFL to a Moore--Read state as $|D|$ increases at $\nu=\frac{9}{2}$, but remains a CFL over the entire displacement-field range at $\nu=\frac{15}{2}$. 
}
\label{fig.theoryfig}
\end{figure}

In contrast, the orbitals of the excited LLs are strongly modified by a displacement field. Specifically, the first excited LLs of $H_0$ occur at energies $E_\pm = \pm \sqrt{2}\hbar \omega_\mathrm{c}$ with eigenstates $(\phi_2,\pm \phi_{0})$. Turning on a displacement field couples these levels via $\langle +|H_D|-\rangle = \frac{\xi u}{2}$ and changes the orbital composition by an amount $\sim  u/\hbar \omega_\mathrm{c}$. In the full four-band model, there are two additional components with $\phi_1$ orbitals whose weight is much smaller and changes only weakly with $D$. The complete orbital weights of all four bands are shown in Fig.~\ref{fig.theoryfig}a and derived in  Sec.~6 in \cite{SM}.

In the excited LLs, the $K$ and $K'$ valleys are affected oppositely, depending on the sign of $D$. For negative $D$, the weight of $\phi_2$ orbitals in the $K$ valley increases at the expense of $\phi_0$, while it decreases for $K'$. The two valleys are degenerate for $H_0+H_D$, but are weakly split by microscopic corrections that are omitted in $H_0$ for clarity \cite{McCann_BLG_2006, McCann_BLG_2013, Jung_Accurate_BLG_2014, hunt_direct_2017}. This opposite change in the orbitals explains the qualitatively different evolution of fractional quantum Hall plateaus at large $D$, where the fillings $4<\nu<6$ realize $K$ and $6<\nu<8$ realize $K'$. 

\textit{Numerical evidence of orbitally-stabilized pairing}---To support the intuition that increasing the $N=2$ admixture favors pairing, we perform exact diagonalization of the half-filled first excited LL. Specifically, we find the single-particle eigenstates of the non-interacting four-band model at different $D$.  We use these orbitals to compute Coulomb matrix elements and to diagonalize the interacting Hamiltonians. Finally, we compare the resulting ground states to two standard trial states: the compressible CFL~\footnote{As a trial state of the CFL phase, we use the ground state of Coulomb in the $N=0$ LL of non-relativistic electrons.} and the incompressible Moore--Read state. To simplify the discussion, we focus on the levels partially occupied at $4\leq \nu\leq5$ and $7\leq \nu\leq8$, which do not exhibit level crossings at $D\neq 0$. 

Our numerical results for 20 electrons in a spherical geometry~\cite{DiagHam} are shown in Fig.~\ref{fig.theoryfig}b, see Sec.~14 in \cite{SM} for calculations with other system sizes. At $\nu=\frac{9}{2}$, the ground state evolves from a CFL into a Moore--Read state with increasing $|D|$. The same results also apply for $\nu=\frac{11}{2}$ at $|D|\gtrsim 230~\mathrm{mV/nm}$. By contrast, the ground state at $\nu=\frac{15}{2}$ (and $\nu=\frac{13}{2}$ at $|D|\gtrsim 230$mV/nm) depends only weakly on the displacement field and has a large overlap with the CFL throughout. In either case, we find no evidence of a different candidate state at half filling, such as the $f$-wave paired state proposed for the third LL of monolayer graphene~\cite{Kim_Even_Denominator_f_wave_2019}.  

\textit{Discussion}---We have observed and characterized two robust even-denominator states in BLG at $\nu=\frac{9}{2}$ and $\frac{11}{2}$. These states emerge gradually and strengthen with increasing $|D|$, which can be explained by a displacement-field-induced change in the LL orbitals. The $N=2$ components increase significantly over $N=0$ at large displacement fields, modifying the composite-fermion interaction. The occurrence of these incompressible states at large $|D|$ and their insensitivity to an in-plane magnetic field imply a single-component state.

Consistently and perhaps surprisingly, our numerical study also finds that increasing the relative weight of the $N=2$ orbital component promotes CF pairing. This orbital is typically associated with electronic crystals~\cite{Pan_Strongly_1999,Lilly_Evidence_1999}, and indeed, we expect such phases to emerge at even larger displacement fields, where the orbital composition approaches a pure $N=2$ level~\cite{Goerbig_Competition_2004,Fogler_Laughlin_wigner_1997}, see Fig.~\ref{fig:triangle_b} in the End Matter. However, accessing such large displacement fields is not feasible due to electric breakdown of the hBN \cite{hattori2015layer,ji2016boron}. At intermediate admixtures, we show that an increase of $N=2$ leads to a similar electron--electron interaction as in pure $N=1$ levels, where CF pairing is well established~\cite{Morf_transition_1998,Park_possibility_1998,Rezayi_incompressible_2000,Scarola_Cooper_2000,Wojs_3body_2005,feiguin_density_2008,Peterson_Orbital_2008,moller_paired_2008,Feiguin_spin_2009,Zaletel_Topological_2013,Sharma_BCS_2021,Sharma_CF_pairing_quart_2024,Yutushui_phase_2025}; see Sec.~6 in \cite{SM}.

Our results suggest that transport measurements in BLG can provide unique insights into the pairing transition of non-Fermi liquids \cite{bonesteel1999,wang2014,Metlitski2015}. In particular, an increase in the gaps at higher magnetic fields may allow the extraction of universal properties near the transition. We note, however, that the magnetic field also changes the orbital composition of the wavefunctions. Even in the zeroth LL, this change could drive transitions between FQH states~\cite{Apalkov_Stable_2011,Balram_BLG_2022,Yutushui_numerical_2025}, but may be overshadowed by the overall change in the interaction scale. Tuning interactions using both $D$ and $B$ as sources of orbital mixing may provide a route for realizing even more exotic FQH phases, such as the non-Abelian Read--Rezayi state \cite{readrezayi1999}.

\section*{Data Availability}
The pseudopotentials and orbital distributions can be generated with the code available at the Zenodo repository~\cite{Zenodo}.
The experimental data that support the findings of this study are available from the corresponding author upon reasonable request.
\section*{acknowledgments}
It is a pleasure to thank Bert Halperin for illuminating discussions. We also gratefully acknowledge support from the Dean of the Faculty and the Clore Foundation (RK); the Quantum Science and Technology Program 2021 and the Schwartz Reisman Collaborative Science Program (YR); the Shimon and Golde Picker–Weizmann Annual Grant (YR); research grants from the Goldfield Family Charitable Trust, the Estate of Hermine Miller, and the Sheba Foundation, Dweck Philanthropies, Inc., and the Tom and Mary Beck Centter for Advanced and Intelligent Materials (C-AIM) grant program (YR); the European Research Council Starting Investigator Grant No.~101163917 (YR); the Minerva Foundation with funding from the Federal German Ministry for Education and Research (DFM and YR); the Israel Science Foundation under grants No.~3281/25 (DFM), No.~380/25 and 425/25 (YR), and No. 3536/26(DFM and YR).

%

\onecolumngrid
\section*{End Matter} \label{app.end_matter}
\twocolumngrid

\begin{figure}[t]
    \centering
    \includegraphics[width=0.8\linewidth]{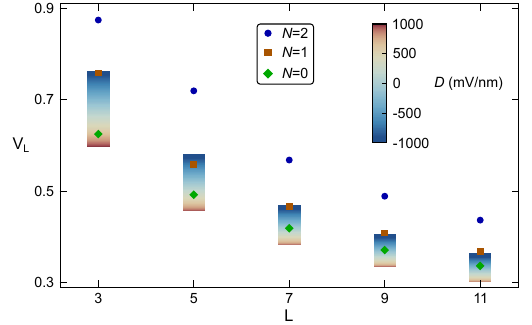}
    \caption{\label{fig:triangle_a}  The Haldane pseudopotentials for the first excited LL of BLG are shown for different displacement fields as indicated on the color bar. The overall normalization is set by fixing $V_{L=1}=1$. For reference, we also show the pseudopotentials for the non-relativistic $N=0,1,2$ LLs. For displacement fields between $-1000\mathrm{mV/nm}$ and $-500\mathrm{mV/nm}$, the pseudopotentials well approximate those of the $N=1$ LL, where pairing is expected to occur.} 
\end{figure}

\textit{Numerical details}---The spinors describing single electron states in the first excited LL of BLG contain $N=0$, $N=1$, and $N=2$ non-relativistic orbitals. Microscopic details, such as the strength of magnetic and displacement fields, strongly modify the precise composition, i.e., the probability amplitudes $C_N$; see Sec.~6 in \cite{SM} for details. The Coulomb interactions between electrons in those states, parametrized by Haldane pseudopotentials $H^\vect{C}_L$~\cite{Apalkov_Stable_2011,Balram_BLG_2022,Yutushui_numerical_2025}, are obtained from their non-relativistic counterparts $H_L$ as 
\begin{align}
    {\cal H}^{\vect{C}}_L = \sum_{N,N'} |C_N|^2|C_{N'}|^2 H_L(N,N').
\end{align}

 \begin{figure}[t]
    \centering
    \includegraphics[width=0.9\linewidth]{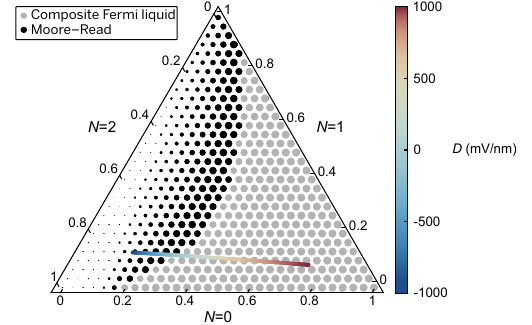}
    \caption{\label{fig:triangle_b} The overlap of Moore-Read and CFL trial states with the ground state of a level composed of $N=0,1,2$ non-relativistic orbitals. The gray or black colors indicate which trial state has the largest overlap, and the marker size represents the magnitude of the overlap. The colored line shows the trajectory of the first excited LL of BLG in the phase space as $D$ is varied (see the color bar). In the low $D$ limit, the orbital composition is composed of the $N=2$ orbital alone, and an additional transition from a Moore-Read state to a Wigner crystal state is expected. } 
\end{figure}

 In a spherical geometry, the Haldane pseudopotentials for a monopole strength $Q$ are given by 
\begin{align}
    H_L(N,N')=(2Q+1)^2 (-1)^{N+N'-L}
 \sum_{k=0}^{2Q}
 \left\{\begin{matrix}2Q-L&Q&Q\\k& Q& Q \end{matrix}\right\}\nonumber
 \\\times
 \begin{pmatrix}Q&k&Q\\N-Q& 0& Q-N \end{pmatrix} 
 \begin{pmatrix}Q&k&Q\\N'-Q& 0& Q-N' \end{pmatrix},\label{eq.Haldane}
\end{align}
where $(\ldots)$ and $\{\ldots\}$ are 3-j and 6-j symbols, respectively. We do not include Landau-level mixing beyond the $D$-dependence of the orbitals. Such additional mixing could arise when strong Coulomb interactions cause electrons to virtually occupy high-energy bands. In the first excited Landau level of BLG, each spin and valley is only represented once, and mixing with levels of identical spin and valley in the zeroth or second LL is suppressed as $E_\text{Coulomb}/\hbar \omega_c$. Consequently, mixing is significantly  weaker than in the zeroth LL, which is composed of two weakly split levels, $n=0$ and $n=1$, for each spin and valley; cf.~Supplementary material of Ref.~\onlinecite{Zibrov_Tunable_bilayer_graphene_2017}

Fig.~\ref{fig:triangle_a} shows how the first five pseudopotentials vary with the displacement field. The amplitudes $C_N$ were computed for $B=12T$ and microscopic parameters listed in SI Sec~6. To facilitate easy comparison, we rescaled the pseudopotentials so that $V_{L=1}=1$. Such a rescaling affects the magnitudes of gaps, but not the competition between different phases. The pseudopotentials of pure $N=0-2$ LLs are shown for reference. For a strong negative displacement field, $H_1$ is suppressed relative to the $H_{L>1}$ pseudopotentials, reflecting a suppression of short-range repulsion. 
In Fig.~\ref{fig.theoryfig}b of the main text, we saw that a strong negative displacement field favors the Moore-Read paired state over a CFL at $\nu=\frac{9}{2}$, based on wave function overlaps. We now extend this analysis to an arbitrary mixture of the $N=0-2$ LLs in Fig.~\ref{fig:triangle_b}. Due to the larger phase space, we use a smaller system with $N_e=12$ electrons. We find that the addition of the $N=2$ component can promote composite Fermion pairing, similarly to the $N=1$ component. As $C_{N=2} \rightarrow 1$, the ground state has vanishing overlap with either trial wavefunction. In this regime, charge density wave states are instead expected to be favorable. 


\begin{table*}[t]
\caption{Studied filling factors and their extracted activation gap and response to in-plane magnetic field}
\begin{ruledtabular}
\begin{tabular}{c|c|c|c|c}
$\nu$ & $D$ range & {$R_{xy}$ value (expected value)($\mathrm{k}\Omega$)} & Activation gap $\Delta$($\mathrm{K}$) & Response to $B_\parallel\lesssim 8$T\\
\hline
$9/2$ & $-540$ to $-1050$ & $5.742~(5.736)$ & $\approx0.1$ (at $D=-1000\mathrm{mV/nm}$) & Insensitive\\
\hline

$11/2$ & $-540$ to $-1050$ & $4.695~(4.693)$ & $\approx0.1$ (at $D=-1000\mathrm{mV/nm}$) & Insensitive\\
\hline

$22/5, 23/5$ & 0 to -900 & $5.870~(5.866),~5.611~(5.612)$ & $\approx0.5$,$0.9$ ($D=-650\mathrm{mV/nm}$) & N/A \\
\hline

$27/5, 28/5$ & 0 to -850 & $4.781~(4.780),~4.612~(4.610)$ & $\approx0.3$,$0.7$ ($D=-650\mathrm{mV/nm}$) & N/A \\
\hline

$13/3, 14/3$ & Entire range & $5.956~(5.957),~5.530~(5.531)$ & $\approx2.6$,$2.3$ ($D=-1050\mathrm{mV/nm}$) & Insensitive\\
\hline

$16/3, 17/3$ & Entire range & $4.838~(4.840),~4.556~(4.555)$ & $\approx2.5$,$2.5$ ($D=-1050\mathrm{mV/nm}$)  & Insensitive\\
\hline

$13/2, 15/2$ & Absent & Compressible & 0 & N/A \\
\end{tabular}
\end{ruledtabular}
\label{tableddata}
\end{table*}

\onecolumngrid
\end{document}